\documentclass[%
reprint,
 amsmath,amssymb,
 aps,
showpacs,showkeys,]{revtex4-1}

\usepackage{graphicx}% Include figure files
\usepackage{dcolumn}% Align table columns on decimal point
\usepackage{bm}% bold math
\begin{document}

%\preprint{APS/123-QED}

%\title{The deuteron wave function and neutron form factors from\\
%the elastic electron deuteron  scattering}% Force line breaks with \\
\title{An algebraic form of the Marchenko inversion. \\
	Partial waves with orbital momentum $l\ge 0$}% Force line breaks with \\
%\title{Energy-independent complex $^{1}S_{0}NN$ potential\\
%from Marchenko equation}% Force line breaks with \\

\author{N. A. Khokhlov}
 %\altaffiliation[Also at ]{Komsomolsk-na-Amure State Technical University}%Lines break automatically or can be forced with \\
\email{nikolakhokhlov@yandex.ru}
\affiliation{%
 Southwest State University, Kursk, Russia
}%
\date{\today}% It is always \today, today,

\begin{abstract}
 We present a generalization of the algebraic method for solving the Marchenko equation (fixed-$l$ inversion) for any values of the orbital angular momentum $l$. 
We expand the Marchenko equation kernel in a separable form using a triangular wave set. 
The separable kernel allows a reduction of the equation to a system of linear equations. 
We obtained a linear expression of the kernel expansion coefficients in terms of the Fourier series coefficients of $q(1-S(q))$ function ($S(q)$ is the scattering matrix) depending on the momentum $q$. 
The linear expression is valid for any orbital angular momentum $l$. 
The kernel expansion coefficients are determined by the scattering data in the finite range $0\leq q\leq\pi/h$. 
In turn, the thus defined Marchenko kernel of the equation allows one to find the potential function of the radial Schrödinger equation with $h$-step accuracy. %The result of the numerical application of the method for $l = 1$ is presented. 
%The corresponding numerical algorithm is obtained for reconstructing complex partial potentials from scattering data on a finite range of $q$. The reconstructed potentials describe with a required accuracy a partial $S$-matrix that is unitary below the threshold of inelasticity and non-unitary (absorptive) above the threshold. 
%The developed procedure is applied to analyze the $^{1}S_{0}NN$ data up to 3 GeV. We show that these data are described by energy-independent complex partial potential. 
\end{abstract}
\maketitle

\section{\label{sec:intro}INTRODUCTION}
 The inverse problem (IP) of quantum scattering is essential for many physical applications. One  of the most important such applications is the interparticle potential extraction from scattering data. 
 The fixed-$l$ IP considered  here is usually solved within the framework of  Marchenko, Krein, and Gelfand-Levitan theories \cite{Gelfand, Agranovich1963, Marchenko1977, Krein1955, Levitan1984, Newton, Chadan}. 
 Development of accurate and unambiguous methods for solving this problem  remains a fundamental challenge 
  \cite{Sparenberg1997,Sparenberg2004,Kukulin2004,Pupasov2011,Mack2012}.
  The ill-posedness of the IP significantly complicates its numerical solution.
 
 This paper considers a new algebraic method for solving the fixed-$l$ inverse problem of quantum scattering theory.  We derive the method from the Marchenko theory. 
   Marchenko theory was successfully applied by H.V. von Geramb and H. Kohlhoff to recover nucleon-nucleon partial potentials from partial-wave analysis (PWA) data up to the inelastic threshold ($E_{\text{lab}}\approx~280$~MeV) \cite{Geramb1994, Kohlhoff1994}.
   They used rational fraction expansions of partial $S$-matrices. This expansion allows one to obtain an analytical solution to the Marchenko equation  (Bargman-type potentials).   Optical model nucleon-nucleon partial potentials were recovered from PWA data up to 3~GeV using a similar approach \cite{Khokhlov2006, Khokhlov2007}.  
   It is not clear whether such a procedure converges with an increase  in the $S$-matrix approximation accuracy. 
   We approximate the integral kernel by a separable series in the triangular  single wave set. Thus, the Marchenko equation is solved analytically as in
Refs.~\cite{Geramb1994, Kohlhoff1994, Khokhlov2006, Khokhlov2007}. 
The expansion coefficients of the integral kernel are obtained from the Fourier series coefficients of the function $q\left(1-S(q)\right)$ on a finite range ($0\leq q\leq\pi/h$) of the momentum $q$. The  $h$  value determines a required accuracy of the potential function. 

\section{Marchenko equation in an algebraic form}
The radial Schrödinger equation is
\begin{equation}
	\label{f1}
	\left(\frac{d^{2}}{d r^{2}}-\frac{l(l+1)}{r^{2}}-V(r)+q^{2}\right) \psi(r, q)=0.
\end{equation}
 The Marchenko equation  \cite{Agranovich1963, Marchenko1977} is a Fredholm integral equation of the second kind:
\begin{equation}
	\label{f3}	F(x, y)+L(x, y)+\int_{x}^{+\infty} L(x, t) F(t, y) d t=0
\end{equation}
The kernel function is defined by the following expression
\begin{multline}
	F(x, y)=\frac{1}{2 \pi} \int_{-\infty}^{+\infty} h_{l}^{+}(q x)[1-S(q)] h_{l}^{+}(q y) d q \\	
	+\sum_{j=1}^{n_{b}} h_{l}^{+}\left(\tilde{q}_{j} x\right) M_{j}^{2} h_{l}^{+}\left(\tilde{q}_{j} y\right)\\
	=\frac{1}{2 \pi} \int_{-\infty}^{+\infty} h_{l}^{+}(q x) Y(q) h_{l}^{+}(q y) d q
	\label{f4}	
\end{multline}
where $h_{l}^{+}(z)$ is the Riccati-Hankel function, and
\begin{equation}
	\label{f5}	Y(q)=\left[1-S(q)-i \sum_{j=1}^{n_{b}} M_{j}^{2}\left(q-\tilde{q}_{j}\right)^{-1}\right],
\end{equation}
Experimental data entering the kernel are
\begin{equation}
		\label{f2}
		\left\{S(q),(0<q<\infty), \tilde{q}_{j}, M_{j}, j=1, \ldots, n\right\},
	\end{equation}
where $S(q)=e^{2 \imath \delta(q)}$  is a scattering matrix dependent on the momentum $q$. The $S$-matrix defines asymptotic behavior at  $r \rightarrow+\infty$ of regular at $r=0$  solutions of Eq.~(\ref{f1}) for $q \geq 0 ;\ \tilde{q}_{j}^{2}=E_{j} \leq 0, E_{j}$  is $j$-th bound state energy ($-\imath \tilde{q}_{j} \geq 0$); $M_{j}$  is $j$-th bound state asymptotic constant.

The potential function of Eq.~(\ref{f1}) is obtained from the solution of Eq.~\ref{f3}
\begin{equation}
	V(r)=-2 \frac{d L(r, r)}{d r} \label{f6}
\end{equation}
 Many methods for solving  Fredholm integral equations use series expansion of the equation kernel. \cite{eprint7,eprint8,eprint9,eprint10,eprint11,eprint12,eprint13,eprint14}. We also use this approach. 

We introduce auxiliary functions: %consider the case $l = 0$, for which $h_{l}^{+}(q x)=e^{\imath q x}$  and
\begin{equation}
\label{f11b}	F_{m}(z)=\frac{1}{2 \pi} \int_{-\infty}^{+\infty}\frac{e^{\imath qz} Y(q)dq}{q^{m}},
\end{equation}
then
\begin{equation}
	\label{Fm_der}	\frac{d^{k}F_{m}(z)}{dz^{k}}=\imath^{k} F_{m-k}(z),\ (k=1,2,\dots, m).
\end{equation}

\begin{figure}[htb]
	% Use the \centerline and \includegraphics commands to insert your figure file:
	\centerline{\includegraphics[width=0.46\textwidth]{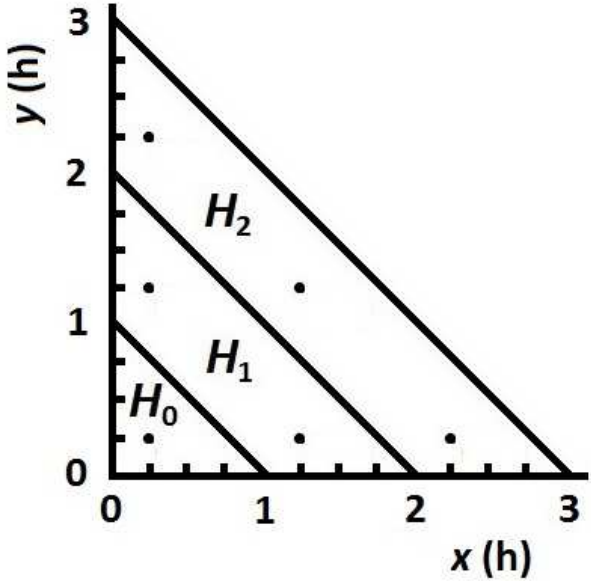}}
	% Use the \caption command to produce the figure caption and place it below the graph:
	\caption{\label{fig:basis1}The basis set $H_{n}\equiv H_{n}(x+y)$ (Eq.~(\ref{rec_set}) is shown as trapezoid (triangle for $n=0$) domains where  $H_{n}(x+y)=1$, and elsewhere $H_{n}(x+y)=0$. The domains are bounded by lines $x=0$, $y=0$, and $x+y=h(n-1)$.
				The basis set $\Delta_{i}(x)\Delta_{j}(y)$ (Eq.~(\ref{tr_set}))  is shown as projections (points) of the corresponding regular square pyramids apexes on the $xy$-plane.
$\Delta_{i}(x)\Delta_{j}(y)=1$ at $x=(0.25+i)h,\ y=(0.25+j)h$ (apex of the $ij$-pyramid).
				 The pyramids bases are $(2h\times 2h)$ squares on the $xy$-plain with sides parallel to the $x$ and $y$ axes. On sides of the corresponding square (as well as outside them)  $\Delta_{i}(x)\Delta_{j}(y)=0$.}
\end{figure}

We use transformations 
\begin{multline}\label{trKrein1}
	\hat{K}_{z,l}f(z)= z^{l+1}\left(-\frac{1}{z}\frac{d}{dz}\right)^{l}\left[z^{-1}f(z)\right]\\
\equiv (-1)^{l} {\sum_{n=0}^{l}\frac{(2l-n)!}{n!(l-n)!}(-2z)^{n-l}\frac{d^{n}f(z)}{dz^{n}} 	}.
\end{multline}
Thus (\cite{Abramowitz}, Eqs.~10.1.23-10.1.26)
\begin{equation}
	\label{trKrein1h1}	
		\hat{K}_{z,l}e^{\pm \imath qz}=q^{l} h^{\pm}_{l}(qz).
\end{equation}
\begin{widetext}
and
\begin{multline}
	\hat{K}_{y,l}	\hat{K}_{x,l}F_{2l}(x+y)
	= {\sum_{n_{1},n_{2}=0}^{l}\frac{(2l-n_{1})!}{n_{1}!(l-n_{1})!} \frac{(2l-n_{2})!}{n_{2}!(l-n_{2})!}
		(-2x)^{n_{1}-l}	(-2y)^{n_{2}-l}\imath^{n_{1}+n_{2}}F_{2l-n_{1}-n_{2}}(x+y)  	} \\
	=\frac{1}{2 \pi} \int_{-\infty}^{+\infty} h_{l}^{+}(q x) Y(q) h_{l}^{+}(q y) d q\equiv F(x, y). \label{FKrein}
\end{multline}
\end{widetext}
%It follows from the  Eqs.~(\ref{trKrein1},\ref{FKrein}) that in order to find $F(x,y)$ we need only %$F_{m}(x+y)$ ($m=2l,2l-1,\dots,0$) 
%
Assuming the finite range $R$ of the bounded potential function, we approximate $F_{m}(x+y)$ as follows:
\begin{eqnarray}
F_{m}(x+y)\approx \sum_{k=-2 N}^{2 N} f_{m,k} H_{k}(x+y) \label{rec_apr}\\
\approx \sum_{k, j=0}^{N} \Delta_{k}(x) f_{m,k+j} \Delta_{j}(y) \label{tr_apr} 
\end{eqnarray}
where $f_{m,k} \equiv F_{m}(k h)$, $h$ is some step, and $R = Nh$. 
The used basis sets are 
\begin{equation}
	\left. \begin{array}{l} H_{0}(x)=\left\{\begin{array}{lr}
			1 & \text{if }  0 \leq x \leq h, \\
			0 & \text{otherwise,}
		\end{array}\right.\\
		H_{n}(x)=H_{0}(x-h n).
	\end{array}\right\} \label{rec_set}
\end{equation}
\begin{equation}
	\left. \begin{array}{l}	\Delta_{0}(x)=\left\{\begin{array}{lr}
					1-|x-0.25| / h & \text{if }  |x-0.25| \leq h, \\
						0 & \text{otherwise,}
		\end{array}\right.\\
		\Delta_{n}(x)=\Delta_{0}(x-h n). 
	\end{array}\right\} \label{tr_set}
\end{equation}
We use bases set $\Delta_{i}(x)\Delta_{j}(y)$ shifted by vector $(0.25h,0.25h)$ compared to the set used previously \cite{MyAlg1,MyAlg2}. %The shift improves convergence.  and avoids singularities on $x$ and $y$ axes for  
The basis sets are illustrated in the (Fig.~\ref{fig:basis1}). 
Decreasing the step $h$, one can approach $F_{m}(x+y)$ arbitrarily close at all points with both sets. Coefficients $f_{m,k}$ are same for both approximations. 

The Fourier transform of the basis set Eq.~\ref{rec_apr} 
\begin{equation}
	\tilde{\mathrm{H}}_{k}(q)=\int_{-\infty}^{\infty} \mathrm{H}_{k}(x) e^{-\imath q x} d x=\frac{\imath\left(e^{-\imath q h}-1\right)}{q e^{\imath q h k}}. \label{f13b}
\end{equation}
The Fourier transform of Eq.~(\ref{f11b}) yields
\begin{equation}
	\frac{Y(q)}{q^{m}}\approx \sum_{k=-2 N}^{2 N} f_{m,k} \tilde{\mathrm{H}}_{k}(q)=\sum_{k=-2 N}^{2 N} f_{m,k}  \frac{\imath\left(e^{-\imath q h}-1\right)}{q e^{\imath q h k}}. \label{f13c}
\end{equation}
We rearrange the last relationship
%\begin{widetext}
\begin{multline}
Y(q)/q^{m-1}= \imath \sum_{k=-2 N}^{2 N} f_{m,k} \left(e^{-\imath q h}-1\right)  e^{-\imath q h k}\\
=\imath \sum_{k=-2 N+1}^{2 N}\left(f_{m,k-1}-f_{m,k}\right) e^{-\imath q h k}
+\imath\left(-f_{m,-2 N}\right) e^{\imath q h 2 N}\\
+\imath\left(f_{m,2 N}\right) e^{-\imath q h(2 N+1)}.
	\label{f13dn}	
\end{multline}
Thus, the left side of the expression is represented as a Fourier series on the interval $-\pi / h \leq q \leq \pi / h$. 
\begin{equation}
	f_{m, k-1}  - f_{m,k}
 = -	\frac{ \imath h}{2\pi} \int_{-\pi / h}^{\pi / h}    Y(q)\frac{ e^{\imath q h  k}d q}{q^{m-1}} 	 \label{f14n}
\end{equation}
%\end{widetext}
 for  $k=-2 N, \ldots, 2 N$. We solve the system~(\ref{f14n}) recursively from  $k=2 N+1$ ($f_{m,2 N+1}=0$) for fixed $m$:
  \begin{multline}
  f_{m,k} 
%  	=	\frac{h}{\pi} \int_{0}^{\pi / h} \left[ \delta_{1\, (-1)^m}  \text{Im}\left( Y(q) 
%  	\sum_{\nu = k+1}^{2N+1} e^{\imath q h  \nu}\right) 
%  	-\imath   \delta_{-1\, (-1)^m}  \text{Re}\left( Y(q)
%  	 	\sum_{\nu = k+1}^{2N+1} e^{\imath q h  \nu}\right) \right] \frac{d q}{q^{m-1}}\\
%%%%%%%%%%%%%%%%%%%%%%%%%%%%%%%%%%%%%%%%%%%%%%  	 	
  	 	=  -	\frac{\imath h}{2\pi} \int_{-\pi / h}^{\pi / h}
  \frac{e^{\imath q h (k+1)}\left(1-e^{\imath q h(2N-k+1)} \right)}{\left(1-e^{\imath q h}\right) {q^{m-1}} } Y(q)  
  	 d q.	 	
  	 	 	 \label{f14nn}
  \end{multline}

The $F(x,y)$ is defined by $f_{m,k}$ $(m=0,1,\dots,2l)$, $k=0,1,\dots,2N$ from Eqs.~(\ref{FKrein}), and (\ref{tr_apr}) as 
	\begin{equation}
	F(x, y) \approx  \sum_{k, j=0}^{N} \Delta_{k}(x) F_{k,j}\Delta_{j}(y), \label{f7nn}
	\end{equation}
where 
\begin{widetext}
\begin{eqnarray}
F_{k,j} ={ {\sum_{n_{1},n_{2}=0}^{l}\frac{(2l-n_{1})!}{n_{1}!(l-n_{1})!} \frac{(2l-n_{2})!}{n_{2}!(l-n_{2})!}
		(-2(k+0.25)h)^{n_{1}-l}	(-2(j+0.25)h)^{n_{2}-l} \imath^{n_{1}+n_{2}} f_{2l-n_{1}-n_{2},k+j}  	}} \\
	=
	%%%%%%%%%%%%%%%%%%%%%%%%%%%%%%%%%%%%%%%%%%%%%%  	 	
 -	\frac{\imath h}{2\pi} \int_{-\pi / h}^{\pi / h}  
	h_{l}^{+}(qkh) \frac{e^{\imath q h}\left(1-e^{\imath q h(2N-k-j+1)} \right)}{1-e^{\imath q h}} Y(q)h_{l}^{+}(qjh)  qd q.	 
	\label{f14}
\end{eqnarray}
\end{widetext}
%We calculate taking into account that  $Y(-q)=Y^{*}(q)$. 

Thus, the range of known scattering data defines value of $h$   and, therefore, the inversion accuracy.

 We solve Eq.~(\ref{f3}) substituting
\begin{equation}
	\label{f9n}	L(x, y) \approx \sum_{j=0}^{N} P_{j}(x) \Delta_{j}(y)
\end{equation}
Substitution of Eqs.~(\ref{f7nn}) and (\ref{f9n}) into Eq.~(\ref{f3}), and linear independence of the basis functions give 
\begin{widetext}
	\begin{equation}
		\sum_{m=0}^{N}\left(\delta_{j\, m}+\sum_{n=0}^{N}\left[\int_{x}^{max((m+0.25)h,(n+0.25)h)} \Delta_{m}(t) \Delta_{n}(t) d t\right] F_{n,j}\right) P_{m}(x)	
		=-\sum_{k=0}^{N} \Delta_{k}(x) F_{k,j}
		\label{f10n}	
	\end{equation}
 We define
	\begin{equation}
		\zeta_{n\, m\, p}=\int_{(p+0.25) h}^{max((m+0.25)h,(n+0.25)h)} \Delta_{m}(t) \Delta_{n}(t) d t	
		=\frac{h}{6}\left(2 \delta_{n\, m}\left(\delta_{n\, p}+2 \eta_{n \geq p+1}\right)
		+\delta_{n\,(m-1)} \eta_{n \geq p}+\delta_{n\,(m+1)} \eta_{m \geq p}\right),
		\label{f10b}	
	\end{equation}
\end{widetext}
where $\delta_{k\, p}$ are the Kronecker symbols  $\delta_{k\, p}$, and
\begin{equation}
\eta_{a}=
\left\{\begin{array}{lr}
		1 & \text{if }  a \text{ is true}, \\
		0 & \text{otherwise,}
	\end{array}\right.
	\label{eta}	
\end{equation} 
Since  $\Delta_{k}(h p) \equiv \delta_{k\, p}$, we finally get a system of equations
\begin{equation}
	\label{f11n}	\sum_{m=0}^{N}\left(\delta_{j\, m}+\sum_{n=p}^{N} \zeta_{n\, m\, p} F_{n,j}\right) P_{p, m}=-F_{p,j},
\end{equation}
for $P_{k}(h (p+0.25)) \equiv P_{p, k}$ $(p,k = 0,\dots,N)$ ($j,p = 0,\dots ,N$).   
  
Solution of Eq.~(\ref{f11n}) gives  $P_{p, k}$. We calculate potential values at points $r = hp$ $(p = 0,\dots, N)$ from Eq.~(\ref{f6}) by some finite difference formula.

\section{Results and Conclusions}
\begin{figure}[htb]
	% Use the \centerline and \includegraphics commands to insert your figure file:
	\centerline{\includegraphics[width=0.46\textwidth]{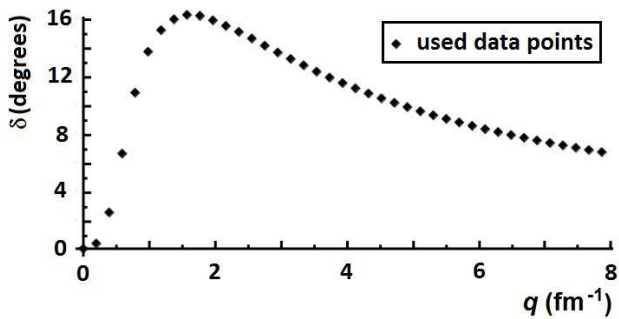}}
	% Use the \caption command to produce the figure caption and place it below the graph:
	\caption{\label{fig:expphase}Data used to reconstruct $V(r)=V_{0}\exp(-ar)$, where $V_{0}=-3\ fm^{-2}= -124.5\ MeV$, $a=1.5\ fm^{-1}$. Units correspond to the $NN$ system.}
\end{figure}

\begin{figure}[htb]
	% Use the \centerline and \includegraphics commands to insert your figure file:
	\centerline{\includegraphics[width=0.46\textwidth]{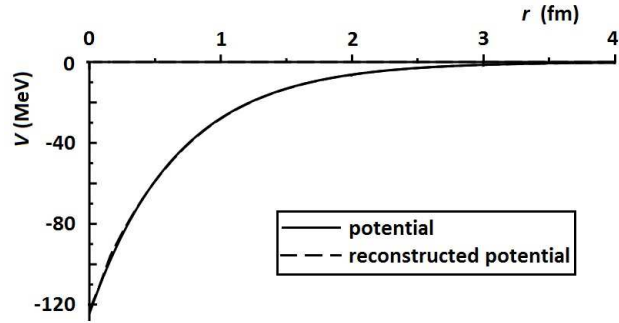}}
	%\centerline{\includegraphics[width=0.8\textwidth]{dwaves.eps}}
	% Use the \caption command to produce the figure caption and place it below the graph:
	\caption{\label{fig:exppot}Initial and reconstructed potentials:  $V(r)=V_{0}\exp(-ar)$, where $V_{0}=-3\ fm^{-2}= -124.5\ MeV$, $a=1.5\ fm^{-1}$. Units correspond to the $NN$ system.}
\end{figure}

We tested the developed approach by restoring the potential function  $V(r)=-3 \exp (-3 r / 2)$ from the corresponding scattering data.    Results are presented in Figs.~\ref{fig:expphase},~\ref{fig:exppot}, where  $h = 0.04$, $R = 4$. 
 The input $S$-matrix was calculated at points shown in Fig.~\ref{fig:expphase} up to $q = 8$.  The $S$-matrix was interpolated by a quadratic spline in the range $0 < q < 8$. The $S$-matrix was approximated as asymptotic $S(q)\approx\exp (-2 i \alpha / q)$  for $q>8$, where  $\alpha$ was calculated at $q = 8$. 

 Thus, we presented a general solution of the quantum scattering inverse problem for the any orbital angular momentum $l$. The algorithm of the solution is as follows. We set the step value $h$, which determines a required accuracy of the potential. From the experimental data, we determine $F_{k,j}$ using Eqs.~(\ref{f14}).
 %for unitary $S$-matrix or from  Eqs.~(\ref{f14b}) for non-unitary $S$-matrix. 
 Solution of Eqs.~(\ref{f11n}) gives values of $P_{k}(hp)$ ($p = 0,\dots,N$). The values of the potential function (\ref{f6}) are determined by some finite difference formula.
Expressions (\ref{f7nn}-\ref{f11n}) give a method for the Marchenko equation's numerical solution for an arbitrary orbital angular momentum $l$.

\end{document}